\documentstyle[prl,aps,amssymb,multicol,twoside,rotate,epsfig]{revtex}

  \makeatletter
  \def\AUTHORS{Technische {Universit\"at} M\"unchen} \def\TITLE{Technische
    {Universit\"at} M\"unchen}

\def\ThisDAY{\the\day\ \ifcase\month{}\or{JANUARY}\or{FEBRUARY}\or{MARCH}%
\or{APRIL}\or{MAY}\or{JUNE}\or{JULY}\or{AUGUST}\or{SEPTEMBER}\or{OCTOBER}%
\or{NOVEMBER}\else{DECEMBER}\fi\ \the\year}

  \def\ps@plain{%
    \gdef\@oddhead{\ifnum\thepage=1 {\hbox to 2in{\sc
          Preprint\hfil}\hfil{{Technische {Universit\"at} M\"unchen}}\hfil\hbox
        to 2in{\hfil\ThisDAY}}\else{\hbox to 1in{\sc
Preprint\hfil}\hfil\TITLE\hfil\hbox to 1in{\hfil\thepage}}\fi}%
\gdef\@evenhead{\hbox to 1in{\thepage\hfil}\hfil\AUTHORS%
\hfil\hbox to 1in{\hfil\sc Preprint}}%
\gdef\@oddfoot{\ifnum\thepage=1 {\hbox to 2.5in{\hfil}\hfil\thepage%
\hfil\hbox to 2.5in{\hfil Typeset using REV\TeX}}\else{}\fi}%
\gdef\@evenfoot{}}

  \makeatother

  \def\leftrule{\hbox to \textwidth{\vrule width3.375in height.5pt\vrule
      width.5pt height8pt\hfil}\par\bigskip\par}
  \def\rightrule{\bigskip\par\hbox to \textwidth{\hfil\vrule width.5pt
      depth8pt\vrule width3.375in height0pt depth.5pt}\par}

\begin{document}

\draft
\tighten


\title{Determination of the universality class of Gadolinium} \author{E.
  Frey$^1$, F. Schwabl$^1$, S. Henneberger$^{1,2}$, O. Hartmann$^3$, R.
  W\"appling$^3$, A.  Kratzer$^2$, and G.M. Kalvius$^2$}

\address{$^1$Institut f\"ur Theoretische Physik and $^2$Institut f\"ur
  Kernphysik und Nukleare Festk\"orperphysik, Technische Universit\"at
  M\"unchen, D-85747 Garching, Germany; $^3$Institute of Physics, Uppsala
  Universitet, S-75121 Uppsala, Sweden}

\maketitle


\begin{abstract}
  We resolve a longstanding puzzle for the static and dynamic critical behavior
  of Gadolinium by a combined theoretical and experimental investigation. It is
  shown that the spin dynamics of a three dimensional ferromagnet with hcp
  lattice structure and a spin-spin interaction given by both exchange and
  dipole-dipole interaction belongs to a new dynamic universality class, model
  J$^*$. Comparing results from mode coupling theory with results from three
  different hyperfine interaction probes we find quantitative agreement. The
  crossover scenario for the wavevector dependence of the hyperfine relaxation
  rate is determined by a subtle interplay between three length scales: the
  correlation length, the dipolar and the uniaxial wave vector.
\end{abstract}

\pacs{PACS numbers: 75.40.Gb, 75.40.Cx, 76.75.+i, 76.80.+y}

\begin{multicols}{2}

\narrowtext

The spin dynamics of simple ferromagnets in the vicinity of their Curie point
$T_c$ are archetypical examples of dynamic critical phenomena near second-order
phase transitions. Much experimental and theoretical effort has been put into
identifying the dynamic universality classes and assigning them to magnetic
substances. Nevertheless, experimental observations on gadolinium
\cite{collins-chowdhury-hohenemser:86,chowdhury-collins-hohenemser:84,%
chowdhury-collins-hohenemser:86,hartmann-etal:90,waeckelgard-etal:86,%
waeckelgard-etal:89,geldart-etal:89} remained a puzzle up to now. Because of
its large localized magnetic moment, and the fact that it is an S-state ion, Gd
should have a very small magnetocrystalline anisotropy and therefore be much
better a model system for an isotropic Heisenberg magnet than either Fe, Ni or
EuO. As a consequence it should belong to the model J dynamic universality
class in the classification scheme of Ref.~\cite{hohenberg-halperin:77}. The
measured static and especially dynamic critical exponents are, however, not
at all compatible with model J. The objective of this paper is to resolve this
longstanding seemingly contradictory situation by a combined theoretical and
experimental study.

Early experimental observations \cite{cable-wollan:68} clearly demonstrate that
Gd has an easy axis which coincides with the hexagonal axis of its hcp lattice.
The origin of such an easy axis cannot be understood from the
magnetocrystalline anisotropy. But, based on a mean field theory
\cite{fujiki:87} it has been argued that a combined effect of the lattice
structure and dipolar interaction favors the c-axis as the easy direction. This
view is supported by measurements \cite{geldart-etal:89} of the c-axis and
basal-plane susceptibility on a single crystal of Gd. It is found that the
basal-plane susceptibility crosses over from a singular behavior to a constant
at a characteristic temperature scale which can be accounted for by dipolar
effects.  The analysis of the static critical behavior \cite{geldart-etal:89}
is, however, complicated by the fact that all experiments are done in the
non-asymptotic regime where superposed crossover lead to complex temperature
dependences.  This may not be easily interpreted in terms of one or the other
universality class. This is even more so, as the static critical exponents for
the various universality classes are of comparable magnitude.

A surprising and yet unexplained observation was made by a measurement of the
critical dynamics using hyperfine methods 
\cite{collins-chowdhury-hohenemser:86,chowdhury-collins-hohenemser:84,%
  chowdhury-collins-hohenemser:86}. The critical exponent $w$ for the
autocorrelation time $\tau_c$, which should scale as $\tau_c \propto
(T-T_c)^{-w}$ in the asymptotic regime, was found to be $w \approx 0.5$.  The
observed value is not consistent with either Heisenberg or Ising models, but
considerably lower.

The purpose of this letter is twofold. First we give a theoretical description
for a spin system with both exchange and dipolar interaction on a hcp lattice
using mode coupling theory.  Next we calculate the relaxation rates observed in
various hyperfine interaction measurements, where we account for the details of
the coupling tensor in each of these methods.  We also report on measurements
of the muon spin relaxation ($\mu$SR) rate in high purity single crystal
samples of Gd. A comparison of the theoretical predictions with these and
earlier \cite{hartmann-etal:90,waeckelgard-etal:89} $\mu$SR measurements as
well as perturbed angular correlation (PAC) and Moessbauer data
\cite{collins-chowdhury-hohenemser:86,chowdhury-collins-hohenemser:84,%
  chowdhury-collins-hohenemser:86} gives a coherent picture of the dynamic and
static critical behavior of Gd and resolves the puzzling situation described
above.

We consider a system with $N$ identical spins fixed on the sites of a three
dimensional lattice. Taking into account magnetocrystalline anisotropy as well
as dipolar interaction it is described by the Hamiltonian
\begin{eqnarray}
H = -\!\sum_{i \neq j} 
\left[ J_{i j}^{\perp} \left( S^x_i \! S^x_j \!+\! S^y_i \! S^y_j \right) 
\!+\! J_{i j}^{\parallel} S^z_i \! S^z_j 
\!+\! D_{ij}^{\alpha \beta} S_i^\alpha \! S_j^\beta \right].
\label{hamiltonian}
\end{eqnarray}
The magnitude of the magnetocrystalline anisotropy of the system is
characterized by $\Delta = J^{\parallel} / J^{\perp}$. The dipolar interaction
is characterized by the tensor
\begin{equation}
D_{ij}^{\alpha \beta} = 
- \frac{(g_L \mu_B)^2}{2} 
\left( \frac{\delta_{\alpha \beta}}{|{\bf x}_{ij}|^3}
     - \frac{3 x_{ij}^\alpha x_{ij}^\beta}{|{\bf x}_{ij}|^5} 
 \right) ,
\end{equation}
with ${\bf x}_{ij} = {\bf x}_i - {\bf x}_j$, $g_L$ is the Land\'e factor, and
$\mu_B$ the Bohr magneton.  Dipolar lattice sums, $D_{\bf q}^{\alpha \beta} =
\sum_{i \neq j} D_{ij}^{\alpha \beta} e^{i {\bf q} \cdot {\bf x}_i }$, can be
evaluated by using Ewald's method. For infinite three--dimensional {\em cubic
  lattices} the results may be found in
Ref.~\cite{aharony-fisher:73,frey-schwabl:94}.  For Bravais lattices with a
{\em hexagonal-closed packed (hcp) structure} the dipolar tensor to leading
order in ${\bf q}$ becomes \cite{fujiki:87}
\begin{eqnarray}
 D_{\bf q}^{\alpha \beta} = \frac{(g_L \mu_B)^2}{2 v_a}
 \left[\beta_4^\alpha \delta_{\alpha \beta} 
      - 4 \pi \frac{q_\alpha q_\beta}{q^2} + {\cal O}(q^2) \right],
\end{eqnarray}
where $v_a$ is the volume of the primitive unit cell with lattice constant $a$,
and the parameters are $\beta_4^x = 4.12$ and $\beta_4^z = 4.32$. Upon
expanding the Fourier transform of the exchange interaction $J_{\bf q}^\alpha =
{\sum_i}^\prime J_{i0}^\alpha e^{i {\bf q} \cdot {\bf x}_i } \approx J_0^\alpha
- J q^2 a^2 + {\cal O}(q^4)$, and keeping only those terms which are relevant
in the spirit of the renormalization group theory, one finds
\begin{eqnarray}
  H = J \sum_{\bf q} 
      \Biggl[ \biggl(m^\alpha  -  \Delta_0^\alpha +  q^2 a^2
              \biggr) \delta_{\alpha \beta}
                     + g  \frac{q_\alpha q_\beta}{q^2} \Biggr]
      S^\alpha_{- \bf q} S^\beta_{\bf q} .
\label{effective_hamiltonian}
\end{eqnarray}
There are two sources of uniaxial anisotropy in the Hamiltonian,
Eq.~\ref{effective_hamiltonian}, magnetocrystalline anisotropy,
$\Delta_0^\alpha = J_0^\alpha/J$, and dipolar interaction, $m^\alpha = (g_L
\mu_B)^2 \beta_4^\alpha / 2J v_a$.  In addition, the dipolar interaction
introduces an anisotropy of the spin-fluctuations with respect to the wave
vector ${\bf q}$ which is reflected by the term proportional to $q_\alpha
q_\beta / q^2$. The magnitude $g$ of this anisotropy is given by $g = 4 \pi
(g_L \mu_B)^2 / 2 J v_a$.  We define a dimensionless quantity $m = (g_L
\mu_B)^2 (\beta_4^\parallel - \beta_4^\perp) / 2 J v_a$ proportional to the
ratio of the anisotropy energy and the exchange energy.  Putting in values for
Gd the ratio of the dipolar contribution to the term $q_\alpha q_\beta / q^2$
and to the uniaxial anisotropy is $\sqrt{g/m} = 7.8738$.  In the following we
will show that all available data for Gd can be explained by assuming that the
uniaxial anisotropy is solely due to the dipolar interaction. Therefore, we
will assume $\Delta = 0$ in the following discussion.

Now we turn to an analysis of the static critical behavior. In Ornstein-Zernike
approximation the static susceptibility reads
\begin{equation}
\chi^{-1}_{\alpha \beta}({\bf q}) = J \left[ (r_{\alpha} + q^2) 
\, \delta_{\alpha \beta} + 
q_D^2 \, \frac{q_{\alpha} q_{\beta}}{q^2} \right],
\label{propagator}
\end{equation}
where $r_z = r = \xi^{-2}$, $r_{x,y} = \xi^{-2} + q_A^2$ and we have measured
all length scales in units of the lattice constant $a$. The analysis of the
critical behavior resulting from the Hamiltonian, Eq.\ 
\ref{effective_hamiltonian}, is complicated by the fact that besides the
correlation length $\xi = \xi_0 (T/T_c -1)^{-\nu}$ there are two anisotropy
length scales $q_A^{-1}=a/\sqrt{m}$ and $q_D^{-1}=a/\sqrt{g}$, both resulting
from the dipolar interaction.  The eigenvalues of the inverse susceptibility
matrix are given by $ \lambda_1({\bf q}) = q^2 + \xi^{-2} + q_A^2$ and
$\lambda_{2,3}({\bf q}) = q^2 + \xi^{-2} + \left[q_D^2 + q_A^2 \pm W \right]/2$
where $W = [(q_D^2+q_A^2)^2-4 q_D^2 q_A^2 q_z^2/q^2]^{1/2}$; the eigenvectors
${\bf e}_i ({\bf q})$ are given in a forthcoming publication
\cite{henneberger-etal:97}.  It is interesting to note that due to the combined
effect of the dipolar interaction and the uniaxial anisotropy of the lattice
the eigenvalues of the susceptibility matrix remain finite in the limit $q
\rightarrow 0$ and upon approaching the critical temperature. Only if the angle
$\vartheta$ between the easy axis ($z$-axis) of magnetization and the wave
vector is $\vartheta = 90^o$ the third eigenvalue becomes critical. Due to the
anisotropy terms in the Hamiltonian there are two crossover as a function of
the reduced temperature and wave vector. To a good approximation these can be
incorporated in an effective exponent of the correlation length $\xi$
\cite{ried-millev-faehnle-kronmueller:96b}.

Mode coupling theory is a theoretical method which has been shown to give
highly accurate results for the critical dynamics of cubic ferromagnets
\cite{frey-schwabl:94}. Here we generalize this method to non-cubic magnets.
Starting from the equations of motion for the components $s^{\alpha}_{\bf q}$
of the spin ${\bf S}_{\bf q}$ in the eigenvector basis, $s^{\alpha}_{\bf q} =
\sum_i S^i_{\bf q} e_{\alpha i}({\bf q})$, one can derive the following set of
coupled integral equations \cite{henneberger-etal:97} for the half-sided
Fourier transform, $\Phi_\alpha ({\bf q}, \omega) = i \chi_{\alpha} ({\bf q}) /
[\omega + i \Gamma_\alpha ({\bf q}, \omega)]$, of the Kubo relaxation function
$\Phi_{\alpha \beta} ({\bf q},t)$,
\begin{eqnarray}
\Gamma_\alpha ({\bf q},\omega)
& = & \frac{4 k_B T J^2}{\chi_{\alpha}({\bf q})} 
\int_{{\bf k},\omega'} \sum_{\beta \gamma}
K^{\beta \gamma }_{\alpha} ({\bf k}, {\bf q}) \nonumber \\
&\times& \Phi_\beta ({\bf k}, \omega)
\Phi_\gamma ({\bf q}\!-\!{\bf k}, \omega\! - \!\omega'), 
\label{mode_coupling_equations}
\end{eqnarray} 
where $\int{{\bf k}, \omega} = \int (d^3 k / (2 \pi)^3) \int (d \omega / 2
\pi)$.  The vertex functions $ K^{\beta \gamma}_{\alpha} ({\bf k}, {\bf q}) $
for the decay of the mode $\alpha$ into the modes $\beta$ and $\gamma$ are
given by \cite{henneberger-etal:97}
\begin{eqnarray}
K^{\beta \beta }_{\alpha} ({\bf k}, {\bf q}) & = &
T^{\beta \beta}_{\alpha} ({\bf k}, {\bf q}) 
U_{\alpha \beta}^{\beta}({\bf k}, {\bf q}) 
\left[ \lambda_{\beta} ({\bf k})  -  
       \lambda_{\beta}({\bf q}-{\bf k}) \right] , \nonumber \\
K^{\beta \gamma }_{\alpha} ({\bf k}, {\bf q}) & = &
T^{\beta \gamma}_{\alpha} ({\bf k}, {\bf q}) \,
T^{\beta \gamma}_{\alpha} ({\bf k}, {\bf q}) ,  \quad \beta \not = \gamma,
\label{vertices}
\end{eqnarray}
with $U_{\alpha \beta}^\gamma({\bf k}, {\bf q}) = \sum_{ijk} \varepsilon_{ijk}
\, e_{\alpha i} ({\bf k}) \, e_{\beta j} ({\bf q}) \, e_{\gamma k} ({\bf
  q}-{\bf k})$, where $\varepsilon_{ijk}$ is the Levi-Cevita symbol, and
$T^{ii}_{\alpha} ({\bf k}, {\bf q}) = \lambda_{i} ({\bf k}) U_{i
  \alpha}^{i}({\bf k}, {\bf q})$, $T^{i j}_{\alpha} ({\bf k}, {\bf q}) = \left[
  \lambda_{i} ({\bf k})- \lambda_{j}({\bf q}\!-\!{\bf k}) \right] U_{i
  \alpha}^{j}({\bf k}, {\bf q})$ for $i<j$ and $T^{i j}_{\alpha} ({\bf k}, {\bf
  q}) = 0$ for $i>j$.  In the limits $q_D \rightarrow 0$ or $q_A \rightarrow 0$
these equations reduce either to the uniaxial or to the isotropic dipolar
ferromagnet \cite{frey-schwabl:94,bagnuls-joukoff_piette:75}

We have solved the above mode coupling equations in the Lorentzian
approximation for the lineshape \cite{henneberger-etal:97}. It is found that
the mode coupling equations obey a generalized dynamic scaling law where the
linewidth depend on three scaling variables, $x_1=1/q\xi$, $x_2=q_D/q$ and
$x_3=q_A/q$, and the angle $\vartheta$ between the direction of the wave vector
and the $z$-axis. The explicit functional form is too complicated to be
presented in this letter and we refer the reader to a forthcoming publication
\cite{henneberger-etal:97}. But, note that it is implicitly contained in the
damping rates for the hyperfine interaction probes discussed below.

\begin{figure}
\centerline{\epsfig{figure=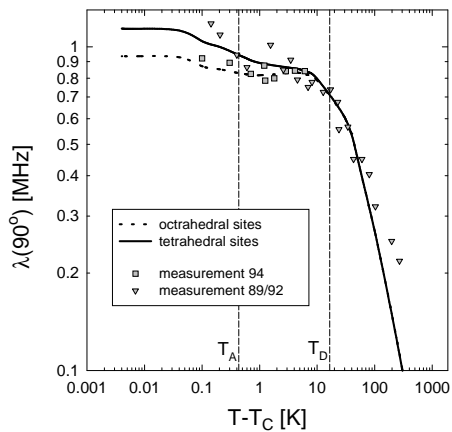,height=6.0cm,clip=}}
\caption{Experimental and theoretical results of the relaxation rate 
  $\lambda$ for tetrahedral and octahedral muon sites with $\alpha = 90^o$.
  Data taken from Ref.~\protect\cite{waeckelgard-etal:89,hartmann-etal:90} and
  measured at the $\mu$SR facility at the PSI in 1992 and 1994 (see inset)
  \protect\cite{henneberger-exp-etal:97}}
\label{fig:90}
\end{figure}

Now we compare the above theoretical results with experimental data obtained
from hyperfine interaction experiments.  We have performed zero field $\mu$SR
experiments on Gd \cite{henneberger-exp-etal:97}.  Spin polarized muons were
implanted in order to measure the distribution and the dynamics of the internal
magnetic field at the muon site via the temporal loss of the initial spin
polarization. The experiments were performed at the $\mu$SR facility of the
Paul Scherrer Institute using a high momentum muon beam.  The sample was a
spherical Gd single crystal with diameter $2.5$ cm. The temperature could be
stabilized to at least $\pm 0.05 K$. We could describe the temporal loss of the
initial muon spin polarization by an exponential decay function $P(t) = \exp
\left( - \lambda_z t \right)$. Taking into account anisotropic dipolar fields
as well as the isotropic Fermi contact field to the local field at the muon
site the muon damping rate can be written as
\cite{yaouanc-reotier-frey:93,reotier-yaouanc:94,henneberger-etal:97}
\begin{equation}
\lambda_{\hat z} = \frac{\pi {\cal D}}{V^2}
\int_{{\bf q}} 
\sum_{{\hat \beta} {\hat \gamma}} 
\left[ 
G^{{\hat x} {\hat \beta}}_{\bf q} G^{{\hat x} {\hat \gamma}}_{-{\bf q}} + 
G^{{\hat y} {\hat \beta}}_{\bf q} G^{{\hat y} {\hat \gamma}}_{-{\bf q}} 
\right]
\Phi^{{\hat \beta} {\hat \gamma}}({\bf q},0) ,
\label{muon_relax_rate}
\end{equation}
where the hatted variables indicate that the corresponding quantities have to
be evaluated in the muon reference frame, i.e.\ the ${\hat z}$-axis coincides
with the initial polarization of the muon beam. Here we have defined ${\cal D}
= \gamma_{\mu}^2 (\mu_0/4 \pi)^2 (g_L \mu_B)^2$.  The coupling of the muon spin
and the spins of the magnet is described in terms of the coupling matrix
$G^{{\hat x} {\hat \beta}}_{\bf q}$, which reflects the particular symmetry of
the lattice sites occupied by the muons. Since the most dominant contribution
to the damping rate comes from wave vectors close to the Brillouin zone center,
the form of the coupling tensor $G^{{\hat x} {\hat \beta}}_{\bf q}$ at small
values of ${\bf q}$ will be important.  In the limit ${\bf q} \to 0$
\cite{reotier-yaouanc:94,henneberger-etal:97} one finds $G^{\alpha \beta}_{{\bf
    q} \to 0} = - 4 \pi \left[ {q_{\alpha} q_{\beta} / q^2} - p_{\alpha}
\right] , $ with $p_{\alpha} = d_{\alpha} + n_{\mu} H_{\mu} / 4 \pi$.  With the
Fermi contact field $B_{\rm FC} = - 6.98 $~kG at $T = 0$~K
\cite{denison-graf:79}, one gets {$n_{\mu} H_{\mu}/4 \pi = -0.278$}
\cite{reotier-yaouanc:94}, and consequently for octahedral sites $p_x = p_y =
0.0705$ and $p_z =0.0250$ \cite{reotier-yaouanc:94}, and for tetrahedral sites
$ p_x = p_y = 0.0338$ and $p_z =0.0984$ \cite{henneberger-etal:97}.  The muon
relaxation rate $\lambda_z$ depends on the material parameters $q_A \xi_0$ and
$q_D \xi_0$. Since we assume that both anisotropies result from the dipolar
interaction the ratio $q_D / q_A = 7.8738$ is known and the number of material
parameters is reduced to one. In comparing our theory with $\mu$SR experiments
at a polarization $\alpha = 90^o$ we get the best fit to the data with $q_D
\xi_0 = 0.13$ (see Fig.~\ref{fig:90}). This results in the following values for
the uniaxial and dipolar wave vector, $q_A = 0.0165 / \xi_0$, $q_D = 0.13 /
\xi_0$. The corresponding crossover temperatures are given by $T_A = T_c +
0.43$~K and $T_D = T_c + 16.54$~K suggesting the following crossover scenario.
For $T \gg T_D$ we expect critical behavior dominated by the (isotropic)
Heisenberg fixed point. The relaxation rate shows power law behavior $ \lambda
\propto t^{-w} , $ with an exponent $w \approx 1$. For temperatures in the
interval $T_D > T > T_A$ dipolar interaction becomes important. But, from the
analysis of the uniaxial crossover \cite{henneberger-etal:97} it turns out that
the uniaxial crossovers in dynamics sets in at wave vectors much larger than
expected from an analysis of the static quantities. Therefore, even for $T >
T_A$ we expect to observe effects from dipolar interaction as well as uniaxial
anisotropy.  Finally, for $T < T_A$ the critical dynamics is determined by the
uniaxial dipolar fixed point. Then the static susceptibilities do no longer
diverge for $q \rightarrow 0$ and $T \rightarrow T_c$ except when the wave
vector is perpendicular to the easy axis of magnetization. Since the relaxation
rate $\lambda_z$ is given by an integral over the whole Brillouin zone, the
relative weight of the critical axis along which the susceptibility diverges
becomes vanishingly small. As a consequence the relaxation rate $\lambda_z$ no
longer diverges for $T \rightarrow T_c$, i.e.\ $w \rightarrow 0$ (compare
Fig.~\ref{fig:90}).

Fig.~\ref{fig:90} shows a comparison between the theoretical and experimental
results for an initial polarization inclined by an angle $\alpha = 90^o$ with
respect to the easy axis of magnetization. The solid line and the dashed line
are the theoretical result for the muon relaxation rate if the muons
penetrating the sample are located at tetrahedral and octahedral interstitial
sites, respectively. The comparison between theory and experiment favors
tetrahedral sites. This is confirmed by $\mu$SR experiments with the initial
polarization along the easy axis of magnetization. The ratio $\lambda_z (90^o)
/ \lambda_z (0^o)$ for $T \rightarrow T_c$ becomes $1.2$ and $0.7$ for
octahedral and tetrahedral sites, respectively \cite{henneberger-exp-etal:97}.
The experiment is closer to the latter value strongly suggesting that muons
occupy octahedral sites within the Gd lattice.

The coupling tensor in PAC and Moessbauer measurements reduces to a Fermi
contact coupling. As a consequence the observed relaxation rate is a sum over
the eigenmodes $\tau_c = (k_B T / 3 v_a) \sum_{\alpha} \int_{\bf q}
\chi_{\alpha}({\bf q}) / \Gamma_{\alpha}({\bf q})$. Important information about
the behavior of the auto-correlation time can be gained from a scaling
analysis. An effective dynamical exponent $z_{\rm eff}$ may defined by $ \tau_c
\propto \left( T-T_c \right)^{-w_{\rm eff}}$ with $w_{\rm eff} = \nu_{\rm eff}
(z_{\rm eff} -1)$, where we have neglected the Fisher exponent $\eta$.  If
dipolar interaction and uniaxial anisotropy were absent, one would expect $w
\approx 1.0$. The dipolar interaction is known to be a relevant perturbation at
the Heisenberg fixed point. It leads to asymptotic static critical exponents
which are only slightly different from the corresponding Heisenberg values.
But, since dipolar interaction implies a non-conserved order parameter the
asymptotic dynamic exponent becomes $z_{\rm D} \approx 2$ resulting in a
crossover from $w_{\rm I} \approx 1.0$ to $w_{\rm D} \approx 0.7$. Uniaxial
interaction is also known to be a relevant perturbation with respect to the
Heisenberg fixed point.  Again, the static critical exponents are not changed
very much, e.g.  one finds the Ising (I) value $\nu_{\rm I} = 0.63$, but the
dynamic exponent becomes $z_{\rm I} \approx 4$ if the order parameter is
conserved ($z_{\rm I} \approx 2$ otherwise). The corresponding exponent for the
hyperfine relaxation rate would turn out to be $w_{\rm I} \approx 1.89$ and
$w_{\rm I} \approx 0.63$ for conserved and non-conserved order parameter,
respectively. According to these scaling arguments it is hard to think of any
dynamic universality class which could lead to an effective exponent $w_{\rm
  eff}$ smaller than about $0.6$.  Actually, however Moessbauer studies and PAC
measurements on Gd show distinctly anomalous low values $w \approx 0.5$, which
cannot be explained by either of the above scenarios. This experimental puzzle
can be resolved if one considers the combined effect of dipolar interaction and
uniaxial anisotropy.  As we have seen in the above analysis of the static
critical behavior of uniaxial dipolar ferromagnets, {\em all} eigenvalues of
the susceptibility matrix remain finite upon approaching the critical
temperature except when the wave vector of the spin fluctuations is
perpendicular to the easy axis. Since this is only a region of measure zero in
the Brillouin zone one actually expects that the relaxation rate does no longer
diverge upon approaching $T_c$, i.e., $w \rightarrow 0$.

\begin{figure}
\centerline{\epsfig{figure=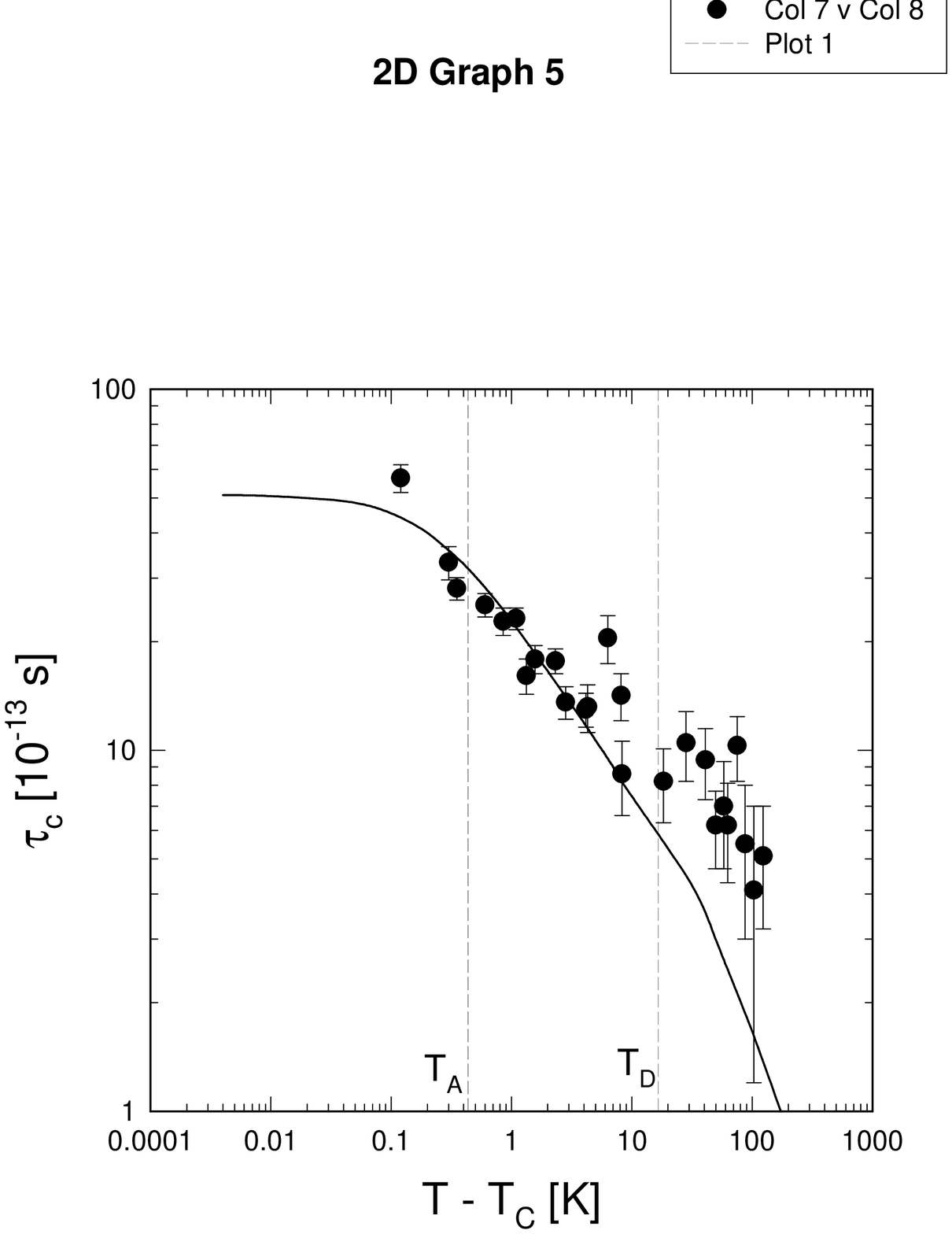,height=6.0cm,clip=}}
\caption{Experimental and theoretical results of the 
  autocorrelation time $\tau_c$ for PAC-experiments. Data taken from\
  \protect\cite{collins-chowdhury-hohenemser:86}.}
\label{fig:pac}
\end{figure}

\begin{figure}
  \centerline{\epsfig{figure=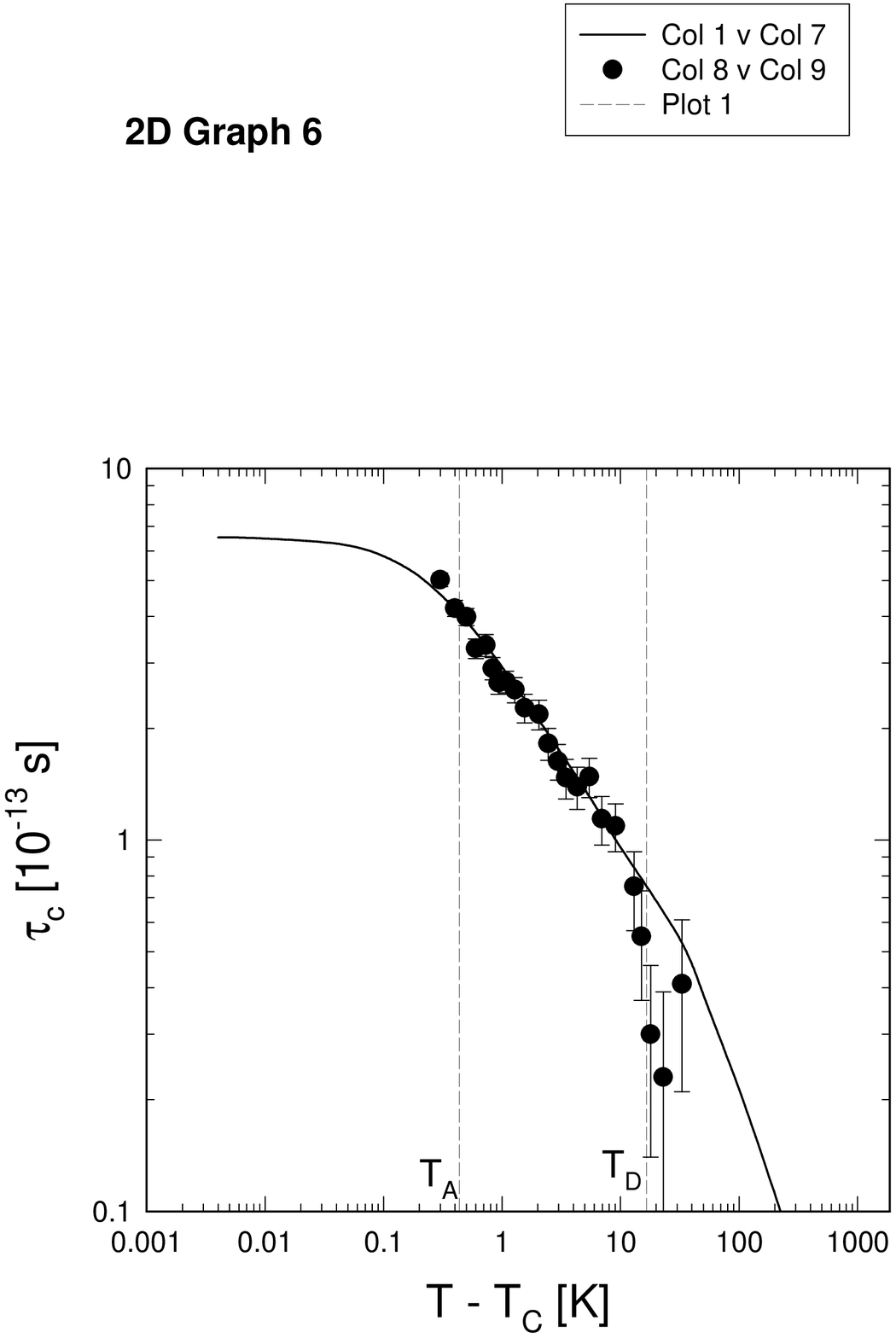,height=6.0cm,clip=}}
\caption{Experimental and theoretical results of the autocorrelation time
  $\tau_c$ for Moessbauer spectroscopy experiments. Data taken from
  \protect\cite{chowdhury-collins-hohenemser:84,chowdhury-collins-hohenemser:86}.}
\label{fig:moess}
\end{figure}

Let us now compare the results of our mode coupling theory with hyperfine
experiments on Gd mentioned above \cite{collins-chowdhury-hohenemser:86,%
  chowdhury-collins-hohenemser:84,chowdhury-collins-hohenemser:86}.  The
autocorrelation time $\tau_c$ is shown in Figs.~\ref{fig:pac} and
\ref{fig:moess} for PAC experiments and Moessbauer spectroscopy, respectively.
Both sets of data are in excellent agreement with the results from mode
coupling theory for $T-T_c < 10K$. Note that besides the overall frequency
scale there is no fit-parameter, since we have used the same set of values for
the dipolar and uniaxial wave vector as for our comparison with $\mu$SR
experiments.

In summary, we have outlined a mode coupling theory for uniaxial dipolar
ferromagnets, where the uniaxiality solely results from the dipolar
interaction. We have also reported measurements on a high purity single crystal
sample of gadolinium metal in the paramagnetic regime.  From the quantitative
agreement between this theory and our $\mu$SR and previous PAC and Moessbauer
measurements the following conclusions can be drawn: (i) The universality class
of Gd is the uniaxial dipolar ferromagnet, where both the isotropic dipolar and
the uniaxial contribution to the spin Hamiltonian are due to a combined effect
of dipolar interaction and non-cubic lattice structure. This corresponds to a
new anisotropic model J$^*$, where the Hamiltonian is given by
Eq.~\ref{effective_hamiltonian}. (ii) The dominant factor for the uniaxial
anisotropy in Gd is the dipolar interaction. (iii) Muons in Gd are located at
octrahedral interstitial sites close to $T_c$. 

{\it Acknowledgment:} It is a pleasure to acknowledge helpful discussions with
A.  Yaouanc and P. Dalmas de R\'eotier. This work has been supported by the
German Federal Ministry for Education and Research under Contract No.~03SC4TUM
and No.~03KA2TUM4. E.F. acknowledges support from the Deutsche
Forschungsgemeinschaft under Contract No.~Fr.~850/2.

\end{multicols} 

\end{document}